\documentclass[aps,prl,twocolumn,showpacs]{revtex4-1}
\usepackage{graphicx}  
\usepackage{subfigure}  
\usepackage{bm}        
\usepackage{amssymb}   
\usepackage[tbtags]{amsmath}
\usepackage{exscale}
\usepackage{xcolor}
\newcommand{\ket}[1]{|#1\rangle}
\newcommand{\bra}[1]{\langle#1|}

\definecolor{darkred}{rgb}{0.90,0.2,0.2}

\usepackage[plainpages=false,pdfpagelabels,colorlinks=true,linkcolor=red,urlcolor=blue,citecolor=blue,pdftitle={Quantum quenches and relaxation dynamics in the thermodynamic limit},pdfauthor={},pdfdisplaydoctitle=true,pdfduplex=DuplexFlipLongEdge]{hyperref}

\begin{document}

\title{Quantum Quenches and Relaxation Dynamics in the Thermodynamic Limit}

\author{Krishnanand Mallayya}
\affiliation{Department of Physics, Pennsylvania State University,
University Park, Pennsylvania, USA 16802}
\author{Marcos Rigol}
\affiliation{Department of Physics, Pennsylvania State University,
University Park, Pennsylvania, USA 16802}


\begin{abstract}
We implement numerical linked cluster expansions (NLCEs) to study dynamics of lattice systems following quantum quenches, and focus on a hard-core boson model in one-dimensional lattices. We find that, in the nonintegrable regime and within the accessible times, local observables exhibit exponential relaxation. We determine the relaxation rate as one departs from the integrable point and show that it scales quadratically with the strength of the integrability breaking perturbation. We compare the NLCE results with those from exact diagonalization calculations on finite chains with periodic boundary conditions, and show that NLCEs are far more accurate.
\end{abstract}

\maketitle

While recent years have seen remarkable advances in understanding when and why isolated quantum systems thermalize after taken far from equilibrium \cite{dalessio_kafri_16, eisert_friesdorf_review_15, polkovnikov2011colloquium}, a fundamental open question is how observables approach their thermal values, and how the equilibration process is affected by approaching an integrable point (in which thermalization does not occur) \cite{moeckel_kehrein_2008, *moeckel_kehrein_2009, rigol2009breakdown, *rigol2009quantum, eckstein_kollar_09, kollar_wolf_11, nessi_iucci_14, bertini2015prethermalization, *bertini2016prethermalization, brandino_caux_15}. Research in this field has been strongly motivated by experiments with ultracold quantum gases, for which near-unitary dynamics can be achieved for exceptionally long times \cite{kinoshita2006quantum, trotzky2012probing, gring2012relaxation, langen_erne_15, kaufman_tai_16, clos_porras_16, tang2017thermalization}. 

Isolated quantum systems are usually taken far from equilibrium using quantum quenches. This means that the initial state (with density matrix $\hat\rho_I$) is a stationary state of an initial time-independent Hamiltonian $\hat{H}_I$, and then, at a time that we denote as $\tau=0$, $\hat{H}_I$ is suddenly changed to a time-independent Hamiltonian $\hat H$ such that $[\hat H, \hat{H}_I]\ne 0$. The resulting unitary evolution of the density matrix $\hat\rho(\tau)=e^{-i\hat H \tau}\hat{\rho}_Ie^{i\hat H \tau}$ generates nontrivial dynamics of observables $\hat O$ (we set $\hbar=1$), whose expectation values are given by $O(\tau)=\text{Tr}[\hat{\rho}(\tau)\hat O]$. One says that an observable has equilibrated when it exhibits small fluctuations about its infinite time average $\bar{O}=\lim_{\tau'\rightarrow\infty}\frac{1}{\tau'}\int_{0}^{\tau'}O(\tau) d\tau$, also known as the prediction of the diagonal ensemble $\bar{O}=O_\text{DE}\equiv\text{Tr}[\hat{\rho}_{\text{DE}}\hat{O}]$, where $\hat{\rho}_{\text{DE}}=\lim_{\tau'\rightarrow\infty}\frac{1}{\tau'}\int_{0}^{\tau'}\hat\rho(\tau) d\tau$ \cite{rigol2008thermalization}. Thermalization occurs when results from the diagonal ensemble and traditional statistical mechanics agree \cite{dalessio_kafri_16}, and is generally a consequence of eigenstate thermalization \cite{rigol2008thermalization, deutsch1991quantum, srednicki1994chaos}.

Theoretical studies of large many-body quantum systems far from equilibration are challenging because of the complexity of quantum dynamics. Full exact diagonalization calculations provide access to the dynamics at arbitrary long times, for arbitrary Hamiltonians and initial states, but are restricted to small system sizes \cite{rigol2008thermalization, rigol2009breakdown, *rigol2009quantum, khatami_pupillo_13, zangara_dente_13, sorg_vidmar_14, tang2017thermalization}. Techniques such as the time-dependent density matrix renormalization group and related approaches \cite{schollwock2005density, *schollwock2011density} and dynamical mean field theory \cite{Aoki14, georges1996dynamical} can be used to study the thermodynamic limit but are limited to the short-time dynamics, and to one and infinite dimensions, respectively. Finally, for weak quenches, Boltzmann and equations of motion techniques have been used at short and intermediate times \cite{stark_kollar_13, lux2014hydrodynamic, bertini2015prethermalization, *bertini2016prethermalization}. 

Here, we introduce numerical linked cluster expansions (NLCEs) as a controllable nonperturbative technique to study dynamics following quenches in lattice systems in the thermodynamic limit. NLCEs can be used to study dynamics of pure and mixed (thermal equilibrium) states in arbitrary dimensions. They have been shown to be a powerful tool to study quantum systems in thermal equilibrium \cite{rigol2006numerical, *rigol2007numerical1, *rigol2007numerical2}, entanglement entropy \cite{kallin_hyatt_13, stoudenmire_gustainis_14, sherman_devakul_16}, diagonal ensemble predictions after quantum quenches from initial thermal \cite{rigol2014quantum, *rigol2016fundamental} and pure \cite{wouters_denardis_14_93, rigol2014quantum2, piroli_vernier_16} states, and for obtaining steady-state results in driven dissipative systems \cite{biella2017linked}.

NLCEs are based on the linked cluster theorem \cite{domb1972phase, guttmann_book_89, oitmaa2006series}, which states that an extensive quantity per site $\mathcal{O}$ can be calculated as the sum of contributions from all connected clusters that can be embedded on the lattice:
\begin{equation}\label{nlce_eq}
\mathcal{O}=\sum_{c}\mathcal{L}(c)\, W_{O}(c),
\end{equation}
where $\mathcal{L}(c)$ is the number of ways per site that cluster $c$ can be embedded on the lattice, and $W_O(c)$ is the weight of the observable in the cluster. $W_O(c)$ is computed using the expectation value of observable $\hat{O}$ in the cluster $O^c=\text{Tr}[\hat{\rho}^c\hat{O}]$, where $\hat{\rho}^c$ is the density matrix of the cluster ($O^c$ is calculated using exact diagonalization), and the inclusion exclusion principle,
\begin{equation}\label{weight_subtraction}
W_{O}(c)=O^c- \sum_{s \subset c} W_{O}(s),
\end{equation}
where the sum runs over all connected subclusters of $c$. For the smallest cluster, $W_{O}(c)=O^c$.

For thermal equilibrium calculations, $\hat{\rho}^c = e^{-(\hat{H}^c-\mu\hat{N}^c)/T} / \text{Tr}[e^{-(\hat{H}^c-\mu\hat{N}^c)/T}]$ is the grand canonical density matrix for the cluster, where $\hat{H}^c$ and $\hat{N}^c$ are the Hamiltonian and the total number of particle operator in the cluster, respectively (we assume that $[\hat{H}^c,\hat{N}^c]=0$), $T$ is the temperature, and $\mu$ is the chemical potential \cite{rigol2006numerical, *rigol2007numerical1, *rigol2007numerical2}. For diagonal ensemble calculations after quenches, $\hat{\rho}^c= \sum_{n^c} (\bra{n^c}\hat{\rho}^c_I\ket{n^c}) \ket{n^c}\bra{n^c}$ is the diagonal ensemble density matrix for the cluster, where $\ket{n^c}$ are the eigenkets of the post-quench Hamiltonian in the cluster (we assume they are nondegenerate), and $\hat{\rho}^c_I$ is the initial density matrix of the cluster \cite{rigol2014quantum, *rigol2016fundamental}. (See Ref.~\cite{tang2013short} for a pedagogical introduction to NLCEs, and Ref.~\cite{mallayya2017numerical} for a detailed discussion of NLCEs for the diagonal ensemble.)

Here, we use NLCEs to study dynamics after quantum quenches in which the initial state is a thermal equilibrium state of the initial Hamiltonian $\hat{H}_I$, i.e., the initial density matrix of each cluster $c$ can be written as $\hat{\rho}_I^c = e^{-(\hat{H}_I^c-\mu_I\hat{N}^c)/T_I} / \text{Tr}[e^{-(\hat{H}_I^c-\mu_I\hat{N}^c)/T_I}]$, where $T_I$ and $\mu_I$ are the initial temperature and chemical potential, respectively. In order to compute expectation values of observables with NLCEs at time $\tau$, one needs to use the density matrix of each cluster $c$ at that time, $\hat{\rho}^c(\tau) = \left(e^{-i\hat{H}^c\tau}\right)\hat{\rho}_I^c\left( e^{i\hat{H}^c\tau}\right)$, so that $\mathcal{O}^c(\tau)=\text{Tr}[\hat{\rho}^c(\tau)\hat{\mathcal{O}}]$. $\mathcal{O}^c(\tau)$ is computed using exact diagonalization. 

NLCEs for quantum dynamics of finite-temperature initial states are much more computationally demanding than NLCEs for the diagonal ensemble which, in turn, are more computationally demanding than NLCEs for thermal equilibrium calculations \cite{rigol2014quantum, *rigol2016fundamental}. The reason is that $\hat{\rho}^c(\tau)$ is a dense matrix even in the basis of the eigenstates of the final Hamiltonian (in that basis, the density matrices of the diagonal and thermal ensembles are diagonal), and $\hat{\rho}^c(\tau)$ needs to be computed at each time of interest ($\hat{\rho}^c$ needs to be computed only once for diagonal and thermal ensemble calculations). As for diagonal ensemble calculations, a significant speed up can be gained if one studies dynamics of simple initial pure states \cite{rigol2014quantum2, piroli_vernier_16}.

We focus on the dynamics of hard-core bosons with nearest (next-nearest) neighbor hoppings $t$ ($t'$) and repulsive interactions $V$ ($V'$) in one-dimensional lattices, as described by the Hamiltonian 
{\setlength\arraycolsep{0.5pt}
	\begin{eqnarray}
	\hat{H}&=&\sum_i \left\lbrace -t\left( \hat{b}^\dagger_i \hat{b}^{}_{i+1} + 
	\textrm{H.c.} \right) 
	+V\left( \hat{n}^{}_i-\dfrac{1}{2}\right)\left( \hat{n}^{}_{i+1}-\dfrac{1}{2}\right) 
	\right.\nonumber\\
	&-&\left.t'\left( \hat{b}^\dagger_i \hat{b}^{}_{i+2} + \textrm{H.c.} \right)  
	+V'\left( \hat{n}^{}_i-\dfrac{1}{2}\right)\left( \hat{n}^{}_{i+2}-\dfrac{1}{2}\right)
	\right\rbrace,\label{Hamilt}
	\end{eqnarray}}\\
where $\hat{b}^\dag_i(\hat{b}^{}_i)$ is the hard-core boson creation (annihilation) operator and $\hat{n}^{}_i=\hat{b}^\dag_i\hat{b}^{}_i$ is the number operator at site $i$. For $t,V\ne 0$, this model is integrable when $t'=V'=0$ and nonintegrable otherwise \cite{cazalilla_citro_review_11}. NLCE results for the diagonal ensemble after quantum quenches within this model were reported in Ref.~\cite{rigol2014quantum, *rigol2016fundamental}. Because of the presence of next-nearest neighbor hoppings and interactions, there is freedom as to the class of clusters that one can use in NLCE studies of this model. This has been discussed Ref.~\cite{mallayya2017numerical}, where the NLCE with maximally connected clusters, namely, clusters with contiguous sites and all bonds present, was shown to be optimal to study the diagonal ensemble. We use those clusters to study the time evolution. The number of sites $l$ of the largest maximally connected cluster considered defines the order of the NLCE result for observables, denoted here by $\mathcal{O}_l(\tau)$.
 
We consider $t_I=0.5$, $V_I=1.5$, and $t'_I=V'_I=0$ for $\hat{H}_I$. After the quench, $t=V=1$ (setting the unit of energy and time) and $t'=V'\in(0,0.8)$. The initial temperature is set to $T_I=10$ (qualitatively similar results are obtained for other temperatures), and we consider systems at half filling. The thermal density matrix $\hat{\rho}_{\text{GE}}$ used to prove thermalization following the quench is characterized by a temperature $T$ and a chemical potential $\mu$, which are set to match the energy, $\text{Tr}[\hat\rho_\text{GE}\hat H]=\text{Tr}[\hat \rho_{I}\hat H]\label{finalT}$, and the number of particles, $\text{Tr}[\hat\rho_\text{GE}\hat N]=\text{Tr}[\hat \rho_{I}\hat N]$, of the initial state after the quench. Particle-hole symmetry in $\hat{H}_I$ and $\hat{H}$ means that $\mu=\mu_I=0$ at half filling. All calculations in the grand canonical ensemble are done through the 19th order of the NLCE, for which all quantities reported are converged within machine precision and are indistinguishable from the exact diagonalization ones in periodic chains with 20 sites~\cite{iyer2015optimization}. 

\begin{figure}[!t]
\includegraphics[width=0.48\textwidth]{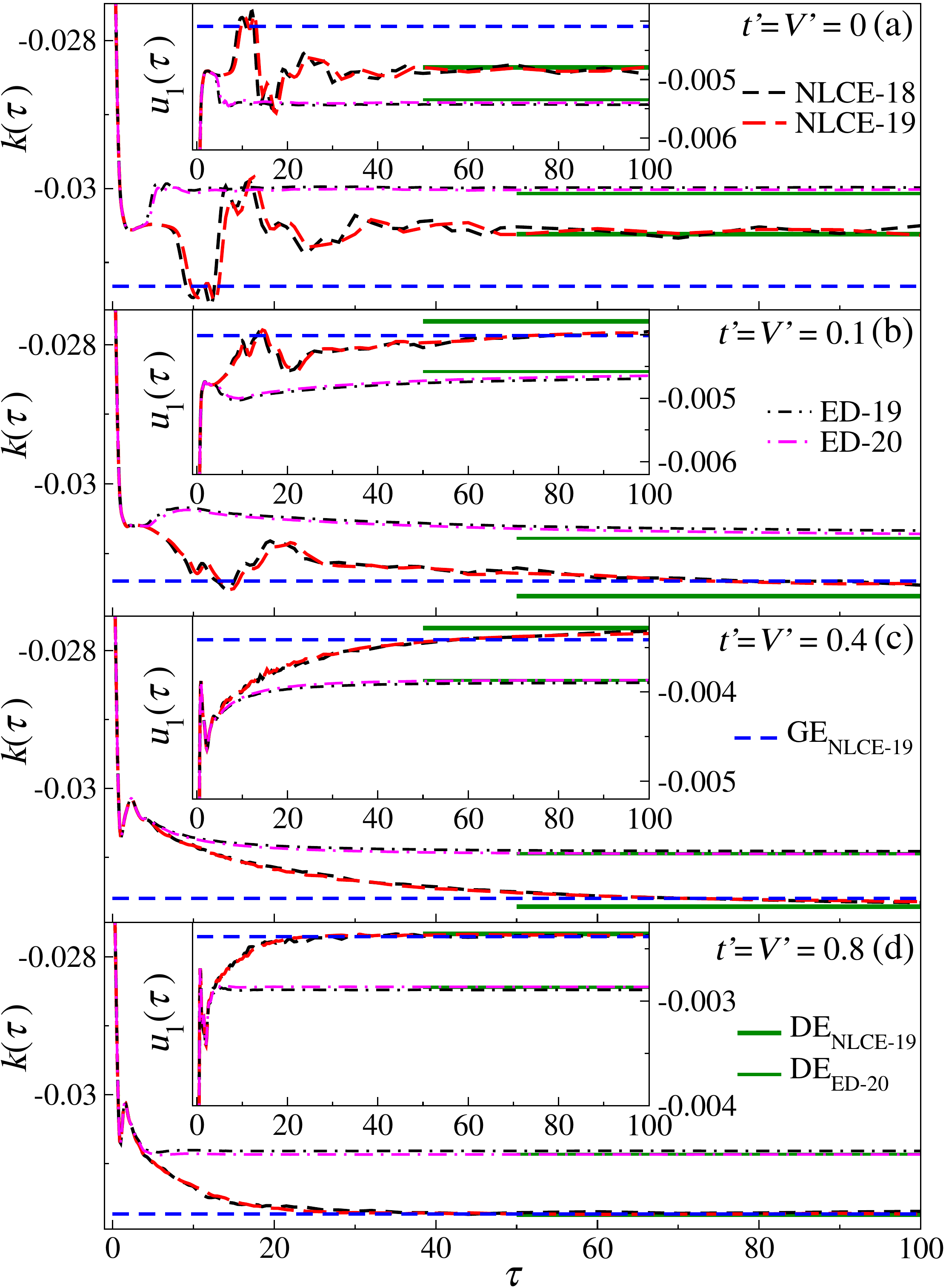}
\caption{Dynamics of the kinetic energy per site $k(\tau)$ (main panels) and of the nearest neighbor interaction energy per site $u_1(\tau)$ (insets) following quantum quenches in which (a) $t'=V'=0$, (b) $t'=V'=0.1$, (c) $t'=V'=0.4$, and (d) $t'=V'=0.8$, after the quench. Results are shown for the 18th and 19th orders of the NLCE, and for periodic chains with 19 and 20 sites from exact diagonalization. The horizontal lines depict the 19th order result of the NLCE for the diagonal (DE) and grand-canonical (GE) ensembles, and for the DE results from exact diagonalization in chains with 20 sites.}\label{figure1}
\end{figure}

We study the dynamics of two local observables: (i) the total kinetic energy $K(\tau)=\text{Tr}[\hat\rho(\tau)\hat{K}]$, where $\hat{K}=\sum_i-\left(\hat{b}_i^{\dagger}\hat{b}_{i+1}^{ }+\text{H.c}\right)-t'\left(\hat{b}_i^{\dagger}\hat{b}_{i+2}^{ }+\text{H.c}\right)$, per site, denoted as $k(\tau)$, and (ii) the nearest neighbor interaction energy $U_1(\tau) = \text{Tr}[\hat\rho(\tau)\hat{U}_1]$, where $\hat{U}_1=\sum_i\left( \hat{n}^{}_i-1/2\right)\left( \hat{n}^{}_{i+1}-1/2\right)$, per site, denoted as $u_1(\tau)$. The calculations of the dynamics are carried out up to the 19th order of the NLCE, and they are compared to results from exact diagonalization on chains with up to 20 sites and periodic boundary conditions. For the times and observables reported here, the NLCE calculation for each quench takes about 1200 CPU hours, while the exact diagonalization calculations take about 600 CPU hours, in Intel Xeon E5-2680 processors.

In Fig.~\ref{figure1}, we show results for the dynamics of $k(\tau)$ (main panels) and of $u_1(\tau)$ (insets) obtained using NLCE and exact diagonalization of periodic chains. The results for both observables are qualitatively similar. At short times ($\tau\lesssim1$), they exhibit a rapid relaxation dynamics present both at integrability and away from integrability, which is captured accurately by the NLCE and exact diagonalization. At integrability, we find that NLCEs are well converged at short and intermediate times, while still equilibrating to the diagonal ensemble at long times. In between, the NLCE results exhibit a large (localized in time) oscillation that moves slowly toward later times as the order of the expansion is increased. The ultimate fate of this feature is not apparent from the results, but it is likely to become less pronounced or disappear altogether. We note that the difference between the grand-canonical and the diagonal ensemble results at integrability is not due to lack of convergence of the NLCE, but because the system does not thermalize \cite{rigol2014quantum, *rigol2016fundamental}. The difference between the NLCE and exact diagonalization results for the diagonal ensemble indicates the magnitude of the finite size effects in the latter. It shows that finite-size effects are large in the scale of the difference between the diagonal and the grand-canonical ensembles within NLCE, which is due to lack of thermalization.

For $t'=V'>0$, at intermediate and long times, the observables in the NLCE calculations can be seen to approach the diagonal ensemble predictions exponentially fast (see also Fig.~\ref{figure3}). A similar behavior can be seen in the exact diagonalization calculations. However, while the NLCE results for the diagonal ensemble become nearly indistinguishable from the grand-canonical ensemble as $t'=V'$ increase, this does not happen in the exact diagonalization calculations. Hence, since thermalization is expected away from integrability (and seen within NLCE), the disagreement between the NLCE and exact diagonalization results for dynamics at intermediate and long times are essentially the result of large finite-size effects in the exact diagonalization calculations. 

\begin{figure}[!t]
\includegraphics[width=0.48\textwidth]{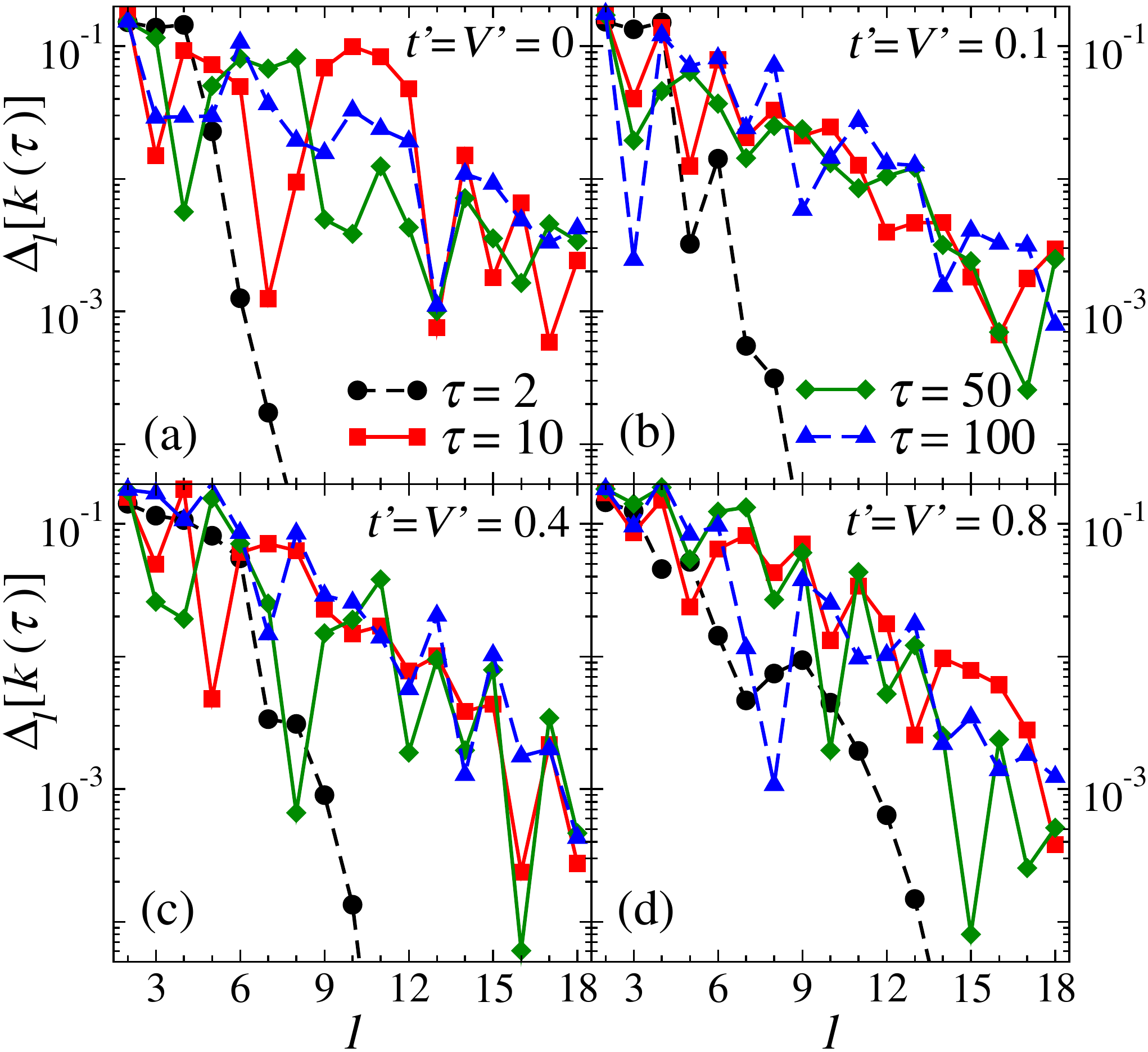}
\caption{Normalized difference $\Delta_l[k(\tau)]$, see Eq.~\eqref{D_relerror}, as a function of $l$ at four times ($\tau=2$, 10, 50, and 100) for (a) $t'=V'=0$, (b) $t'=V'=0.1$, (c) $t'=V'=0.4$, and (d) $t'=V'=0.8$.}\label{figure2}
\end{figure}

Next, we discuss the convergence of NLCE with increasing the order of the expansion, and with increasing the value of $t'=V'$. On the latter, it is already apparent in Fig.~\ref{figure1} that the convergence of NLCE improves as one moves away from integrability. This is because the results for the 18th and 19th order for the dynamics become indistinguishable from each other, and from the thermal prediction at long times. We focus on the kinetic energy per-site (qualitatively and quantitatively similar results were obtained for the other local observables studied). Given the large finite-size effects observed in the exact diagonalization calculations, we do not discuss their convergence here.

To check the convergence of NLCE for $k(\tau)$, we calculate the normalized difference $\Delta_l[k(\tau)]$ between the result at order $l$ and the highest order available to us ($l=19$)
\begin{equation}\label{D_relerror}
\Delta_l[k(\tau)]=\left|\frac{k_{l}(\tau)-k_{19}(\tau)}{k_{19}(\tau)}\right|.
\end{equation}

\begin{figure}[!t]
\includegraphics[width=0.48\textwidth]{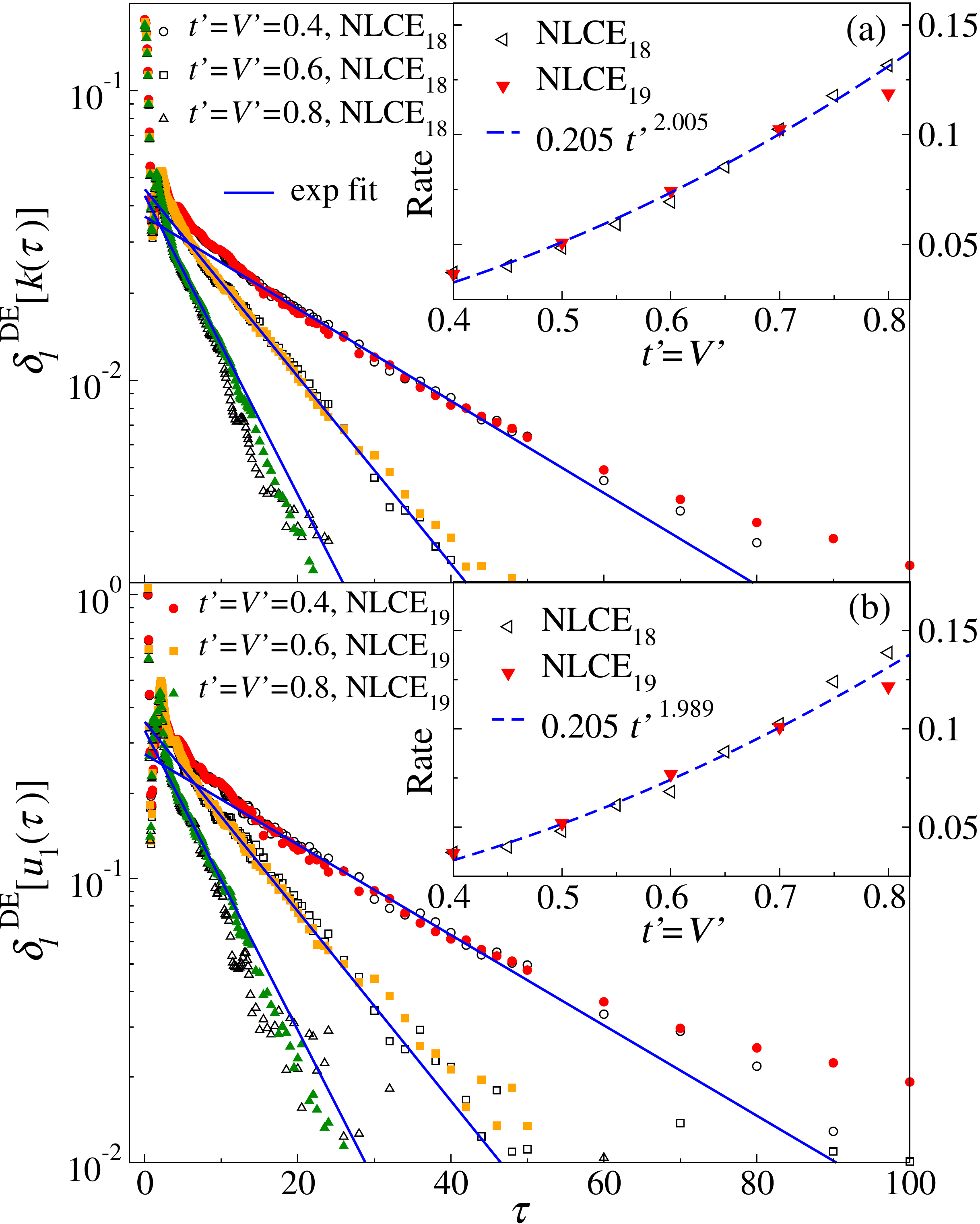}
\caption{(Main panels) $\delta^\text{DE}_l[k(\tau)]$ (a) and $\delta^\text{DE}_l[u_1(\tau)]$ (b) [see Eq.~\eqref{d_relax}] as a function of $\tau$ for $l=18$ (open symbols) and $l=19$ (filled symbols), and for $t'=V'=0.4$, 0.6, and 0.8. The straight lines depict exponential fits to the $l=19$ results, from which we extract the relaxation rate for every value of $t'=V'$. (Insets) Relaxation rates, obtained from fits as those depicted in the main panels, for $l=18$ and $l=19$. The dashed lines depict fits to $a\,t'^{\,\alpha}$ for $0.4\leq t'=V'\leq 0.7$, with $a$ and $\alpha$ reported in the labels of the curves.}\label{figure3}
\end{figure}

In Fig.~\ref{figure2}, we plot $\Delta_l[k(\tau)]$ vs $l$ at four times and for four values of $t'=V'$. At the shortest time ($\tau=2$), $\Delta_l[k(2)]$ can be seen to decrease very rapidly (faster than exponential) with increasing $l$. For the other times depicted ($\tau=10$, 50, and 100), the behavior of $\Delta_l[k(\tau)]$ vs $l$ is consistent with an exponential decay. Comparing Figs.~\ref{figure2}(a) and~\ref{figure2}(b) with Figs.~\ref{figure2}(c) and~\ref{figure2}(d) at the highest order of the NLCE for those times, one can see that $\Delta_l[k(\tau)]$ is about an order of magnitude larger at the integrable point ($t'=V'=0$) and close to it ($t'=V'=0.1$) than far away from integrability ($t'=V'=0.4$ and 0.8). This is similar to findings in Refs.~\cite{rigol2014quantum, *rigol2016fundamental, mallayya2017numerical} for the convergence of diagonal ensemble calculations. The lack of eigenstate thermalization results in a slower convergence for integrable points (there is a large dispersion of eigenstate expectation values of observables in close-by eigenstates), and the closer the system is to integrability the larger the cluster sizes required for the system to ``realize'' that it is nonintegrable (the matrix elements of the integrability breaking perturbation need to become larger than the energy difference between the coupled integrable eigenstates).

Having clarified how the NLCE behaves with increasing the order of the expansion for the observables of interest, we focus in what follows on the relaxation of those observables toward the diagonal ensemble results. In the main panels of Fig.~\ref{figure3}, we plot $\delta^\text{DE}_l[O(\tau)]$, defined as
\begin{equation}\label{d_relax}
\delta^\text{DE}_l[O(\tau)]=\left|\frac{O_{l}(\tau)-O^\text{DE}_{l}}{O^\text{DE}_{l}}\right|,
\end{equation}
for the kinetic energy per site $k(\tau)$ [Fig.~\ref{figure3}(a)] and for the interaction energy per site $u_1(\tau)$ [Fig.~\ref{figure3}(b)]. We report results for $t'=V'=0.4$, 0.6, and 0.8, for $l=18$ and $l=19$. First, we note that the results for $l=18$ and $l=19$ (for both observables) are very close to each other for the values of $t'=V'$ reported (this is less so for smaller values of $t'=V'$, because of the slower convergence close to integrability discussed in the context of Fig.~\ref{figure2}). Second, one can see a regime in time in which $\delta^\text{DE}_l[O(\tau)]$ decays exponentially. In the plots, we report fits to exponentials that make those regimes more apparent. From fits like these, we extract the relaxation rates reported in the insets for $l=18$ and $l=19$. Relaxation rates are only reported for $0.4\leq t'=V'\leq 0.8$ because identifying the exponential decay becomes challenging for $t'=V'\lesssim0.4$, and for $t'=V'\gtrsim0.8$ the decay is so fast that the window in time in which one can fit an exponential becomes very small.

The insets in Fig.~\ref{figure3} show that the results for the relaxation rates are very close for $l=18$ and $l=19$, so that convergence errors appear to be small. More importantly, in agreement with analytic arguments for weakly interacting models \cite{stark_kollar_13, bertini2015prethermalization, *bertini2016prethermalization}, we find that the relaxation rate scales with the square of the strength of the integrability breaking term (see fits in the insets in Fig.~\ref{figure3}). The same scaling was found for the other initial temperatures studied, so this behavior is robust and NLCEs allow one to compute the prefactor accurately in the thermodynamic limit. This is in contrast to exact diagonalization calculations, which suffer from strong finite size effects.

In summary, we introduced NLCEs as a controllable nonperturbative technique to study dynamics following quantum quenches in lattice systems in the thermodynamic limit. NLCEs can be used in arbitrary dimensions, and in this work we focused on a one-dimensional system with nearest and next-nearest neighbor couplings, and on initial thermal equilibrium states. Our NLCE results for different orders exhibit a behavior that is consistent with an exponential convergence independently of the time chosen. We used NLCEs to compute the relaxation rates of two local observables, finding that they scale with the square of the strength of the integrability breaking terms. We should stress that, in contrast to recent theoretical works that studied relaxation rates close to a noninteracting limit \cite{stark_kollar_13, bertini2015prethermalization, *bertini2016prethermalization}, our integrable limit is strongly correlated. Our finding of a quadratic scaling of the relaxation rate with the strength of the integrability breaking perturbation in a strongly interacting system, together with a similar finding in a recent experiment with a strongly interacting dipolar gas \cite{tang2017thermalization}, hints that this behavior is universal close to integrability.

{\it Note added.}---We have become aware of concurrent implementations of NLCEs to study quantum dynamics \cite{whiteNLCE,bakrNLCE}.

\begin{acknowledgements}
This work was supported by the National Science Foundation, Grant No.~PHY-1707482. The computations were carried out at the Institute for CyberScience at Penn State.
\end{acknowledgements}

\bibliographystyle{apsrev4-1}
\bibliography{Reference}

\end{document}